\documentclass[aps,groupedaddress,prd,preprint]{revtex4}
\usepackage{graphicx}
\usepackage[dvips]{pstcol}
\usepackage{pst-node}
\usepackage{pst-plot}
\usepackage{pst-coil}
\begin{document}
\draft
\title{Lepton Flavor Changing Neutral Current Processes at Lepton-Hadron Colliders}
\author{A.T.ALAN and A.SENOL}
\address{Department of Physics, Faculty of Sciences and Arts, Abant Izzet Baysal University,14280 G\"{o}lk\"{o}y, Bolu, TURKEY}
\date{\today}
\begin{abstract}

We investigate the potentials of high energy lepton-proton
colliders to detect a flavor changing $l-l'-Z$ or $l-l'-\gamma$
couplings from the processes ep$\rightarrow \mu(\tau)$X and
$\mu$p$\rightarrow\tau$X. For the $l-l'-Z$ coupling we consider
string inspired E$_{6}$ model whereas $l-l'-\gamma$ coupling is
analyzed in the framework of anomalous magnetic moment type
interaction (AMMI) model.

\end{abstract}
\maketitle

\section{Introduction}
In the Standard Model (SM) lepton flavor is conserved in neutral
current processes at tree level. Many searches for lepton flavor
changing neutral current (FCNC) processes have been performed
within different extensions  of the SM [1]. In this paper we study
e - $\mu$ , e - $\tau$ transitions in ep and $\mu$ - $\tau$
transitions in $\mu$p collisions. For numerical evaluations the
center of mass energy and luminosity values [2] of future
lepton-hadron colliders are given in Tables I and II . The two
dynamical models we use allowing the above transitions are string
inspired E$_{6}$ [3] and anomalous magnetic moment interactions
AMMI [4]. The processes ep$\rightarrow \mu(\tau)$X and
$\mu$p$\rightarrow$$\tau$X occur in t-channel exchange of a
neutral vector boson, Z in E$_{6}$ and $\gamma$ in AMMI .

In E$_{6}$, flavor violating $l-l'-Z$ vertices arise from the
mixings between right handed components of the ordinary and new
heavy chaged leptons:
\begin{equation}\centering\label{1}
  \Gamma^{\mu}=ig_{Z}b_{ll'Z}\gamma^{\mu}(1+\gamma_{5})
\end{equation}
where $g_{Z}=g_{e}/\sin\theta_{W}\cos\theta_{W}$ , $\theta_{W}$ is
 the weak mixing angle and b$_{ll'Z}$ denotes some combination of
the leptonic mixing angles.

In the second model the interaction vertices $l-l'-\gamma$ are :
\begin{equation}\label{2}\centering
  \Gamma^{\mu}=i\kappa_{ll'\gamma}\sigma_{\mu\nu}q^{\nu}
\end{equation}
where q is the momentum transfer through the intermediate photon
and $\kappa$$_{ll'\gamma}$=a$_{ll'\gamma}(\frac{e}{2m_{e}})$
denotes the anomalous magnetic transition moment.
\section{Analysis}
Differential cross section for the subprocess $\l q\rightarrow l'q
$ in E$_{6}$ is obtained as :
\begin{center}
\begin{equation}\label{3}
  \frac{d\hat{\sigma}}{d\hat{t}}=
  \frac{b_{ll'Z}^{2}g_{Z}^{4}}{16\pi
  \hat{s}^{2}[(\hat{t}-M_{Z}^{2})^{2}+M_{Z}^{2}\Gamma_{Z}^{2}]}[(a_{q}+v_{q})^{2}\hat{t}^{2}+
  (a_{q}+v_{q})^{2}(2\hat{s}-m_{l'}^{2})\hat{t}
  +2\hat{s}(a_{q}^{2}+v_{q}^{2})(\hat{s}-m_{l'}^{2})]
\end{equation}
\end{center}
where $\hat{s}=x\mathrm{S}$ is the center of mass energy of the
incoming lepton and quark.
 Using the experimental limits for FCNC Z decays;
 BR(Z$\rightarrow$ e$\mu$) $<$ 1.7$\times$ 10$^{-6}$, BR(Z$\rightarrow$ e$\tau$) $<$ 9.8$\times$ 10$^{-6}$ and BR(Z$\rightarrow$ $\mu$$\tau$) $<$ 1.2$\times$ 10$^{-5}$
 [5] we
obtain the upper bound values of b$_{ll'Z}$:
\begin{center}
\begin{tabular}{c}

  b$_{e\mu Z}<0.504\times10^{-2}$ \\
  b$_{e\tau Z}<1.209\times10^{-2}$ \\
  b$_{\mu\tau Z}<1.338\times10^{-2}$ \\

\end{tabular}
\end{center}
The total cross section is obtained by folding $\hat{\sigma}$ over
the parton distribution functions inside the proton [6] as follow
:
\begin{equation}\label{4}
  \sigma=\int_{x_{min}}^{1}dx\int_{t_{-}}^{t_{+}}{\frac{d\hat{\sigma}}{d\hat{t}}f_{q}(x)}d\hat{t}
\end{equation}
where $x_{min}=\frac{m_{l'}^2}{S}$ and $t_{+}=0,
t_{-}=-(\hat{s}-m_{l'}^{2})$. These phase space boundaries are
obtained in the $m_{e}=m_{q}=0$ case. The results of the
integrated cross sections for various lepton-proton colliders are
presented in Table I. We see that $\sigma(ep\rightarrow\tau
X)\cong 6\sigma(ep\rightarrow\mu X)$.

The differential cross section in AMMI model is calculated as;

 \begin{equation}\label{5}
  \frac{d\hat{\sigma}}{d\hat{t}}=\frac{\alpha
  e_{0}^{2}\kappa_{ll'\gamma}^{2}}{2\hat{s}^{2}}[{m_{l'}^{2}-2\hat{s}+(2\hat{s}m_{l'}^{2}-m_{l'}^{4}-2\hat{s}^{2})\frac{1}{\hat{t}}}]
\end{equation}
where $\alpha$ is the fine-structure constant and $e_{0}$ is the
quark charge in units of the electron charge e. In the integral
(4) we take a $t_{cut}=0.01$GeV to avoid a divergency due to the
photon propagator.
 Now we use the experimental upper bounds for the flavor-violating $\mu$ and
$\tau$ decays; BR($\mu$$\rightarrow$ e$\gamma$) $<$ 1.2$\times$
10$^{-11}$, BR($\tau$$\rightarrow$ e$\gamma$) $<$ 2.7$\times$
10$^{-6}$ and BR($\tau$$\rightarrow$ $\mu$$\gamma$) $<$
1.1$\times$ 10$^{-6}$ to put the limits on a$_{ll'\gamma}$;
\begin{center}
\begin{tabular}{c}
 a$_{\mu e\gamma}<6.36\times10^{-13}$ \\
  a$_{\tau e\gamma}<1.24\times10^{-8}$ \\
  a$_{\tau\mu\gamma}<0.79\times10^{-8}$ \\
\end{tabular}
\end{center}

The results for the integrated cross-sections are given in Table
II.

\begin{center}
\begin{tabular}{|c||c|c|c|c|c|} \hline
  colliders & $\sqrt{S}$(GeV) & $\mathcal{L}$(cm$^{-2}$s$^{-1}$) & $\sigma$(ep$\rightarrow$$\mu$X)(pb) & $\sigma$(ep$\rightarrow$$\tau$X)(pb) &
  $\sigma$($\mu$p$\rightarrow$$\tau$X)(pb)
  \\ \hline\hline
  Thera 1 & 1000 & 4$\times$10$^{30}$ & $0.96\times10^{-2}$ & $5.57\times10^{-2}$ & - \\\hline
  Thera 2 & 1000 & 2.5$\times$10$^{31}$ & $0.96\times10^{-2}$ &$5.57\times10^{-2}$ & - \\\hline
  Thera 3 & 1600 & 1.6$\times$10$^{31}$ & $1.12\times10^{-2}$ & $6.50\times10^{-2}$ & - \\\hline
  LINAC$\otimes$LHC & 5300 & 10$^{33}$ & $1.36\times10^{-2}$ & $7.85\times10^{-2}$& - \\\hline
  $\mu$-Tevatron & 894 & 10$^{31}$ & - & - &$6.48\times10^{-2}$ \\\hline
  $\mu$p & 3000 & 10$^{32}$ & - & - & $9.02\times10^{-2}$ \\ \hline
\end{tabular}
\end{center}
\vspace{0.3cm}
 \begin{center}
 Table I:Cross section values for the FCNC processes
in E$_{6}$
\end{center}
 \vspace{1cm}

\begin{center}
\begin{tabular}{|c||c|c|c|c|c|} \hline
  colliders & $\sqrt{S}$(GeV) & $\mathcal{L}$(cm$^{-2}$s$^{-1}$) & $\sigma$(ep$\rightarrow$$\mu$X)(pb) & $\sigma$(ep$\rightarrow$$\tau$X)(pb) &
  $\sigma$($\mu$p$\rightarrow$$\tau$X)(pb)
  \\ \hline\hline
  Thera 1 & 1000 & 4$\times$10$^{30}$ & 1.71$\times$10$^{-18}$ & 6.54$\times$10$^{-10}$ & - \\\hline
  Thera 2 & 1000 & 2.5$\times$10$^{31}$ & 1.71$\times$10$^{-18}$ &6.54$\times$10$^{-10}$ & - \\\hline
  Thera 3 & 1600 & 1.6$\times$10$^{31}$ & 1.81$\times$10$^{-18}$ & 6.95$\times$10$^{-10}$ & - \\\hline
  LINAC$\otimes$LHC & 5300 & 10$^{33}$ & 2.09$\times$10$^{-18}$ & 8.02$\times$10$^{-10}$ & - \\\hline
  $\mu$-Tevatron & 894 & 10$^{31}$ & - & - & 3.00$\times$10$^{-10}$ \\\hline
  $\mu$p & 3000 & 10$^{32}$ & - & - & 2.65$\times$10$^{-10}$ \\ \hline
\end{tabular}
\end{center}
 \vspace{0.3cm}
 \begin{center}
 Table II:Cross section values for the LFV processes
in AMMI
\end{center}

\section{CONCLUSION}
Our main results indicate that future lepton-hadron colliders are
promising machines for investigation of FCNC processes predicted
by $E_{6}$ model. Unfortunately this statement is not applicable
for LFV processes predicted by AMMI model. According to Table I we
expect $N(e\rightarrow\tau)=785$ and $N(e\rightarrow\mu)=136$
events per year at LINAC$\otimes$LHC and
$N(\mu\rightarrow\tau)=90$ events per year at $\mu$p collider. In
the case of no observations of these events, our results show that
about 2 order of higher accuracy compering to the current limits
will be achieved.

\section{ACKNOWLEDGEMENTS}
We would like to thank S. Sultansoy for useful discussions. This
work is supported in part by Abant Izzet Baysal University
Research Found.

\end{document}